# Direct Imaging of Temperature Evolution of Polar Nanoregions and Chemically Ordered Regions in PMN Relaxor: Evidence for Polar Phase Percolation

Kohei Hino[1], Daisuke Morikawa[2]*, Desheng Fu[3], Mitsuru Itoh[4] and Kenji Tsuda[1,2]

[1]Frontier Research Institute for Interdisciplinary Sciences, Tohoku University, Miyagi 980-0578, Japan

[2]Institute of Multidisciplinary Research for Advanced Materials, Tohoku University, Miyagi 980-8577, Japan

[3]Department of Electronics and Materials Science, Faculty of Engineering and Department of Engineering, Graduate School of Integrated Science and Technology, Shizuoka University, Hamamatsu 432-8561, Japan; Department of Optoelectronics and Nanostructure Science, Graduate School of Science and Technology, Shizuoka University, Hamamatsu 432-8011, Japan

[4]Institute of Science Tokyo, Office of Campus Management, Yokohama 226-8501, Japan

* Author to whom correspondence should be addressed. Electronic mail: daisuke.morikawa.e5@tohoku.ac.jp

**Abstract**

Polar nanoregions (PNRs) are central to understanding the exceptional dielectric and piezoelectric properties of relaxor ferroelectrics and are key to advancing dielectrics for high-energy storage. However, direct real-space imaging of their formation and evolution remains a major challenge in condensed matter physics. Here, we report the real-space mappings of both PNRs and chemically ordered regions (CORs) in the prototypical relaxor $Pb(Mg_{1/3}Nb_{2/3})O_3$ (PMN) and their temperature dependence using convergent-beam electron diffraction (CBED) combined with four-dimensional scanning transmission electron microscopy (4D-STEM). The results reveal that CORs, with sizes of 2–5 nm, remain static with temperature and act to suppress PNR growth. In contrast, PNRs evolve from isolated 2–5 nm regions at room temperature to interconnected structures ~10 nm in size at low temperatures, indicative of a percolation transition. These observations support the random-field model, in which PNRs emerge from a paraelectric matrix and their growth and collective interactions are constrained by random local fields associated with CORs.

Relaxor ferroelectrics exhibit extraordinarily high dielectric permittivity and large piezoelectric responses over a broad temperature range, far surpassing conventional ferroelectrics like $BaTiO_3$[1-4]. These exceptional properties make them indispensable for applications in capacitors, actuators, and transducers. Despite decades of intense experimental and theoretical research, the origin of the "diffuse phase transitions" and frequency-dependent dielectric/piezoelectric responses in these materials remains incompletely understood.

Among relaxors, $Pb(Mg_{1/3}Nb_{2/3})O_3$ (PMN) is the prototypical compound and a central subject in the quest to understand relaxor behavior. PMN adopts a nominally cubic $ABO_3$ perovskite-type structure, with $Pb^{2+}$ occupying the *A*-site and a random distribution of $Mg^{2+}$ and $Nb^{5+}$ on the *B*-site. This chemical and valence disorder on the *B*-site induces nanoscale heterogeneity, giving rise to three distinct local structural components as schematically shown in Figs. 1(a)-(c): (i) a non-polar matrix (NPM) with average $\boldsymbol{Pm\bar{3}m}$ symmetry, (ii) chemically ordered regions (CORs) with $\boldsymbol{Fm\bar{3}m}$ symmetry forming 2 × 2 × 2 superlattice cells due to *B*-site cation ordering, and (iii) polar nanoregions (PNRs), characterized by local displacements of *B*-site cations along ⟨111⟩ directions, associated with $\boldsymbol{R3m}$ symmetry [5-9].

PNRs are considered central to relaxor behavior. They first appear near the Burns temperature ($T_B \approx 600$ K) [1,10] and grow in size with decreasing temperature, reaching 2–5 nm at room temperature. Around the freezing temperature ($T_f \approx 220$ K), a percolation transition is thought to occur, where PNRs rapidly grow (~10 nm) and begin to interconnect [9,11]. In contrast, CORs are believed to remain static in size (~2–5 nm) and distribution, acting as pinning centers that hinder but do not eliminate PNR growth [5-7,9].

Two competing models have been proposed to describe the nature of PNRs and their interactions. The dipole glass model assumes that PNRs are weakly interacting with dipolar entities embedded in a non-polar matrix. Their interactions are random, leading to a frozen, glassy state below a freezing temperature $T_f \approx 220$ K, without the emergence of long-range ferroelectric order [2,12-15]. In this model, the system remains ergodic above $T_f$ and becomes nonergodic below, resulting in a dipole glass phase composed of randomly oriented and frozen polar clusters.

In contrast, the random field model (or ferroelectric nanodomain model) suggests that the system tends intrinsically toward ferroelectric ordering [16-18], but this is frustrated by quenched random electric fields arising from intrinsic *B*-site valence disorder and chemical inhomogeneity in $Pb(B',B'')O_3$ complex perovskites, as demonstrated by theoretical calculations [19-20]. This model predicts that PNRs grow and interact cooperatively, ultimately percolating into nanodomains with ferroelectric correlations at low temperatures. Recent STEM-based polarization mapping studies support this model, revealing nanoscale regions with continuous, albeit frustrated, polarization [21].

While both models acknowledge the existence of PNRs and CORs, the critical distinction lies in the interaction among PNRs and their ground-state behavior. The dipole glass model posits randomly interacting, isolated PNRs that freeze into a disordered state, while the random field model predicts strongly interacting, coalescing PNRs that evolve into a quasi-ferroelectric state at low temperatures. Discriminating between these scenarios requires direct, real-space observation of PNRs and their interactions, which has remained experimentally elusive due to the nanometric size and dynamic nature of these regions.

To overcome these limitations, four-dimensional scanning transmission electron microscopy (4D-STEM) was combined with convergent-beam electron diffraction (CBED). This powerful approach directly visualized the nanometer-scale spatial distribution and temperature evolution of PNRs and CORs in PMN as schematically shown in Fig. 1(d). Specifically, CBED enabled the detection of local symmetry breaking associated with polarization, while 4D-STEM allowed for real-space mapping of such symmetry variations across wide fields of view [22-25]. Through the analysis of two-fold rotational symmetry breaking in CBED patterns and superlattice intensities from CORs, the independent spatial distributions of PNRs and CORs were directly imaged at both room and cryogenic temperatures, providing a means to distinguish between competing theoretical scenarios. Furthermore, by tracking the evolution of these nanostructures across the transition temperature, a critical experimental test of the dipole glass and random field models was achieved. These advanced 4D-STEM-CBED techniques enabled the separate mapping of polar and chemically ordered domains, offering a definitive view of their interplay.

The 4D-STEM experiments were carried out using an omega-type transmission electron microscope (JEM-2010FEF, JEOL) equipped with an in-column energy filter, operated at 100 kV accelerating voltage. Single crystals of $Pb(Mg_{1/3}Nb_{2/3})O_3$ (PMN) were grown using the PbO flux method [11], then mechanically crushed into fine fragments and bonded onto copper TEM grids for observation.

The local thickness of the regions used for data acquisition was estimated to be ~20 nm by comparing experimental CBED patterns to simulations. This thickness was optimized: in thicker regions (>20 nm), background noise hampers the detection of superlattice reflections from chemically ordered regions (CORs), while in thinner areas (<20 nm), the dynamical scattering becomes too weak to resolve subtle symmetry breaking due to polar displacements. Within the specimen thickness of ~20 nm, multiple PNRs and/or CORs may overlap along the electron-beam propagation direction. However, because CBED is highly sensitive to symmetry breaking integrated along the beam path, the observed CBED patterns predominantly reflect the net local symmetry. Since NPM, PNRs, and CORs exhibit distinct symmetry signatures, they can be reliably distinguished even in the presence of partial overlap. CBED patterns were recorded at each probe position using a Gatan STEM diffraction system coupled with a CMOS camera (Gatan RIO16). Two temperature conditions—295 K (room temperature) and 100 K—were employed using a double-tilt liquid nitrogen cooling holder, allowing investigation of structural evolution across the freezing temperature, $T_f \approx 220$ K [9,11].

The scan grid was defined as 50 × 50 points across a 20 × 20 nm$^2$ field of view, using a probe step of 0.4 nm. Each CBED pattern was acquired with a 0.1 s exposure time. Experimental CBED patterns were analyzed by comparing them with simulated patterns computed via *MBFIT* software, which uses Bloch-wave dynamical diffraction theory to account for multiple scattering and enables robust symmetry identification at nanometer scales [26].

Figures 2(a)–(f) illustrate the $[110]_c$ projected crystal structures of the non-polar matrix (NPM) and polar nanoregions (PNRs), along with their corresponding simulated convergent-beam electron diffraction (CBED) patterns. In the NPM, *B*-site atoms ($Mg^{2+}$ and $Nb^{5+}$) occupy their ideal perovskite positions with no off-centering, resulting in a centrosymmetric $\boldsymbol{Pm\bar{3}m}$ structure. The simulated CBED pattern for this configuration exhibits $\boldsymbol{2mm}$ symmetry, characterized by a two-fold rotational axis and two orthogonal mirror planes [Fig. 2(b)].

In contrast, PNRs are modeled with *B*-site atoms displaced along $\langle 111 \rangle$ directions, consistent with an $\boldsymbol{R3m}$ symmetry and local polarization. These displacements were parameterized based on prior structural studies of related relaxors [27]. Due to the cubic symmetry of the parent structure, eight symmetry-equivalent polarization directions are allowed: $\langle 111 \rangle$, $\langle \bar{1}11 \rangle$, $\langle 1\bar{1}1 \rangle$, $\langle 11\bar{1} \rangle$, and their inverses. Two representative projected structures of PNRs are shown in Figures 2(c) and 2(e), with corresponding simulated CBED patterns in Figures 2(d) and 2(f). In both cases, the two-fold rotational symmetry is clearly broken, serving as a signature of local polarization.

Figures 2(g)–(i) present experimental CBED patterns obtained at specific positions marked in the PNR map. The pattern in Figure 2(g), taken from a low-index region, retains two-fold symmetry and is consistent with the NPM. In contrast, the patterns in Figures 2(h) and 2(i) show clear symmetry breaking, confirming their assignment to PNRs. The distinct symmetry types correspond to different $\langle 111 \rangle$ polarization directions, as exemplified in the simulations [Figs. 2(d), 2(f)].

To visualize PNR distributions across the specimen, a **symmetry-breaking index** was calculated for each probe position based on the two-fold rotational symmetry of the CBED patterns [28]. The index is defined as:

$$\text{Symmetry Breaking Index [\%]} = \frac{\sqrt{\sum (I_0 - I_R)^2}}{\sum I_0} \times 100$$

where $I_0$ is the experimental CBED pattern and $I_R$ is its counterpart rotated by 180°. Higher values of the index indicate stronger deviation from two-fold symmetry and thus the presence of local polarization. The resulting real-space map is shown in Figure 2(j), where bright regions correspond to PNRs. Red arrows denote the inferred local polarization directions, determined by matching experimental and simulated CBED disk intensities. The observed directions exhibit slight spatial fluctuations, indicating local disorder or domain variations.

Figures 3(a)–(d) focus on distinguishing chemically ordered regions (CORs) from NPM by analyzing superlattice reflections. Figures 3(a) and 3(c) show the projected *B*-site configurations of NPM and COR, respectively. In the NPM, $Mg^{2+}$ and $Nb^{5+}$ cations are randomly distributed in a 1:2 ratio, consistent with the nominal $Pb(Mg_{1/3}Nb_{2/3})O_3$ composition, resulting in no superlattice reflections in the simulated CBED pattern [Fig. 3(b)]. In contrast, CORs exhibit short-range *B*-site ordering, typically described by $Pb(B'_{1/2}B''_{1/2})O_3$ ($B'$=$Mg_{2/3}Nb_{1/3}$ or Mg, $B''$=Nb), forming a 2 × 2 × 2 superlattice with $\boldsymbol{Fm\bar{3}m}$ symmetry. This ordering produces additional diffraction discs, as shown in the simulated CBED pattern in Fig. 3(d), where a superlattice reflection is marked by a red circle. Two models are commonly proposed for the COR structure: (1) a charge-neutral model [29-30], $Pb[(Mg_{2/3}Nb_{1/3})_{1/2}Nb_{1/2}]O_3$, maintaining the same average composition as bulk PMN; and (2) an ideal 1:1 Mg:Nb ordering model, often referred to as a charge-imbalanced or space-charge model [6-7, 31], $Pb(Mg_{1/2}Nb_{1/2})O_3$, which would require a compensating Nb-rich boundary layer to preserve overall charge neutrality and stoichiometry of the crystal. Support for the charge-imbalanced model has mainly been inferred from donor-doped PMN systems [31]; however, it is not fully consistent with results from resonant x-ray scattering [30], atomic-resolution Z-contrast imaging [32], or COR coarsening observed after annealing in PMN-based relaxors [29], and there is no direct evidence for the expected chemical segregation [33]. In the present work, simulations based on both models yield comparable superlattice reflections in CBED (Fig. S1), making them difficult to distinguish experimentally. Given this limitation and the broader experimental evidence favoring charge neutrality, we adopt the charge-neutral model for the analysis presented here.

The experimental CBED patterns in Figures 3(e) and 3(f), corresponding to probe positions marked on the COR map [Fig. 3(g)], show the spatial variation of superlattice intensities. Figure 3(f) shows clear superlattice reflections, consistent with a COR, while Figure 3(e) lacks such features and exhibits two-fold symmetry, characteristic of NPM. The COR map in Figure 3(g) was generated by evaluating superlattice intensities across the scan area. The color scale spans from 7.2 $e^6$ to 2.5 $e^7$ counts, with brighter regions indicating stronger chemical ordering.

The 4D-STEM–CBED approach enabled us to successfully visualize the spatial distribution of polar nanoregions (PNRs) and chemically ordered regions (CORs) in PMN with sub-nanometer resolution. At room temperature (295 K), CORs exhibit sizes ranging from ~2 nm to 5 nm, in good agreement with previous dark-field TEM observations using $F$-type superlattice reflections, which reported an average size of ~6 nm [5], as well as high-resolution electron microscopy results indicating that CORs are composed of clusters approximately 2 nm in diameter [6].

Simultaneously, PNRs at room temperature are found to be ~5 nm in size, consistent with values inferred from neutron elastic diffuse scattering studies, which estimated a correlation length of ~2 nm [10]. Crucially, the PNR and COR maps presented in Figures 2(j)/4(a) and 3(g)/4(b), respectively, were extracted from the same 4D-STEM dataset, allowing for direct spatial correlation. As illustrated in Figure 4(c), the distributions of PNRs and CORs are spatially distinct and do not overlap, supporting the view that CORs act as local barriers that inhibit—but do not initiate—polarization formation.

Having established the spatial distribution of PNRs and CORs at room temperature, we next examined their evolution upon cooling the crystal to 100 K—below the nominal Curie or freezing temperature (≈ 220 K). This analysis provides a critical test to differentiate between the dipole glass model, which predicts frozen, disordered PNRs at low temperatures, and the random field model, which predicts the cooperative growth and merging of PNRs into nanodomain ferroelectric states.

Figures 4(d) and 4(e) show the PNR and COR maps at 100 K, while Figures 4(a) and 4(b) (replicated from earlier results) serve as direct comparisons at 295 K. Schematic illustrations based on these real-space maps are provided in Figures 4(c) and 4(f). Although minor image drift during acquisition caused a vertical shift and slight elongation of the CORs, comparison of the COR maps at both temperatures confirms that the probe positions remained within nearly the same sample region.

As temperature decreases from 295 K to 100 K, CORs show no appreciable change in either size or distribution, reaffirming their static character. Although slight intensity variations of some CORs are observed between 295 K and 100 K [Figs. 4(b) and 4(e)], these changes do not indicate real structural evolution of CORs. Instead, they are attributed to minor deviations in local diffraction conditions (Bragg conditions) induced by lattice distortions associated with the growth of adjacent PNRs. In stark contrast, PNRs grow significantly, expanding laterally and merging into interconnected networks at 100 K. This percolative behavior suggests the development of a spatially extended, albeit frustrated, polar phase. While complete long-range ferroelectric ordering is prevented by the presence of nearby CORs acting as random-field pinning centers, the observed nanodomain merging is fundamentally incompatible with the dipole glass scenario, which predicts non-interacting, frozen dipoles below freezing temperature $T_f \approx 220$ K. The other results obtained by the different specimen area are shown in Supplemental Materials.

Our results strongly support the random field model of relaxor behavior, in which PNRs—originating from a $\boldsymbol{Pm\bar{3}m}$ paraelectric matrix—grow, interact, and percolate to form a ferroelectric nanodomain state at low temperatures. These findings also corroborate previous reports that the ferroelectric soft mode in PMN recovers below $T_f \approx 220$ K, becoming underdamped and exhibiting a squared frequency that increases linearly with decreasing temperature, consistent with normal ferroelectrics [34]. Thus, our study resolves a long-standing controversy regarding the nature of the ground state in PMN and provides compelling evidence for a percolation-driven transition to a ferroelectric nanodomain state [9,11].

Previously, the nature of PNR polarization and shape has been inferred mainly from neutron and x-ray diffuse scattering, leading to the proposal of a "pancake-shaped" PNR model based on studies of $Pb(Zn_{1/3}Nb_{2/3})O_3$ (PZN) [35]. In contrast, our real-space 4D-STEM–CBED observations show that PNRs in PMN do not exhibit a regular pancake-like morphology, but instead form irregular three-dimensional nanodomains that grow and interconnect upon cooling. It is worth noting that diffuse scattering in relaxor ferroelectrics does not necessarily originate from planar polar regions and can be reasonably explained by oriented domain walls of three-dimensional polar nanodomains, as demonstrated by theoretical simulations [20].

Moreover, while previous local-structure studies based on neutron PDF [8] and EXAFS [36-37] have provided valuable information on short-range correlations, they do not directly resolve the spatial distribution and connectivity of distinct local phases. By contrast, the high sensitivity of 4D-STEM–CBED to local symmetry breaking enables direct real-space mapping of coexisting local structures in PMN, revealing a cubic nonpolar matrix with $\boldsymbol{Pm\bar{3}m}$ symmetry, chemically ordered regions with $\boldsymbol{Fm\bar{3}m}$ symmetry, and polar nanoregions with $\boldsymbol{R3m}$ symmetry. Taken together, the present 4D-STEM–CBED results provide a direct real-space physics picture of the complex local structures in PMN and their temperature evolution. By mapping local symmetry with nanometer resolution, we directly visualize how polar nanoregions (PNRs) emerge from the $\boldsymbol{Pm\bar{3}m}$ paraelectric matrix and progressively expand and interconnect upon cooling, while chemically ordered regions (CORs), with characteristic sizes of ~2–5 nm, remain spatially localized and largely temperature independent. This simultaneous real-space mapping of PNRs and CORs within the same specimen region reveals their distinct roles and mutual interplay, offering direct insight into how nanoscale structural heterogeneity governs the evolution of polar order in relaxor ferroelectrics.

In summary, by employing 4D-STEM in combination with CBED, we have directly visualized the spatial distribution and temperature evolution of polar nanoregions (PNRs) and chemically ordered regions (CORs) in PMN relaxor ferroelectrics with nanometer-scale resolution. Our observations show that CORs remain static with temperature and act to restrict the growth of PNRs, while PNRs grow significantly upon cooling and exhibit percolative connectivity at low temperatures. These results provide direct experimental support for the random field model and definitively rule out the dipole glass scenario. Our findings establish that PMN undergoes a temperature-driven transition into a ferroelectric nanodomain state near 220 K, offering insight into the microscopic mechanisms governing relaxor behavior and paving the way for the design of next-generation dielectric and piezoelectric materials.

See the supplementary material for a comparison of CBED simulations using two different COR models and the temperature evolution of COR and PNR maps obtained from two additional sample areas.

Acknowledgements: We would like to thank Mr. Hideyuki Magara for the careful maintenance of the JEM-2010FEF transmission electron microscope. This work is partly supported by JSPS KAKENHI Grant No. JP22H01150, JP22H04495 JP23K17826 and JP25K01652, the Research Program of Dynamic Alliance for Open Innovation Bridging Human, Environment and Materials in Network Joint Research

Center for Materials and Devices, The Murata Science Foundation, The Mitsubishi Foundation, Yamada Foundation, the Collaborative Research Project of Materials and Structures Laboratory, Institute of Integrated Research, Institute of Science Tokyo and JST PRESTO JPMJPR22AB, Japan.

Data availability: The data that support the findings of this study are available from the corresponding author upon reasonable request.

FIG 1. Schematic illustrations of local structures in PMN: (a) non-polar matrix (NPM), (b) polar nanoregion (PNR), and (c) chemically ordered region (COR). All structures are based on the perovskite-type framework with Pb occupying the *A*-site and Mg/Nb distributed on the *B*-site within oxygen octahedra. In both NPM and PNR, the *B*-site is randomly occupied by Mg and Nb in a 1:2 ratio. Only in PNRs do the *B*-site atoms exhibit off-center displacements along $\langle 111 \rangle$, giving rise to a local polarization vector $P_s$ and $\boldsymbol{R3m}$ symmetry. In contrast, *B*-site cation ordering in CORs involves a (2/3 Mg + 1/3 Nb) occupancy on one site and Nb on another, resulting in an overall 1:2 Mg:Nb ratio. This forms a 2 × 2 × 2 superlattice with $\boldsymbol{Fm\overline{3}m}$ symmetry but lacking spontaneous polarization. (d) Schematic illustration of the 4D-STEM-CBED experiment. A convergent electron probe is raster-scanned across the specimen in a 2D grid. At each probe position, a full CBED pattern is recorded, generating a 4D dataset (2D spatial × 2D diffraction space). The PNRs are identified by analyzing the breakdown of two-fold rotational symmetry in the CBED patterns, while CORs are visualized via the detection of superlattice reflections. This method allows for real-space mapping of local polarization and chemical order with nanometer resolution.

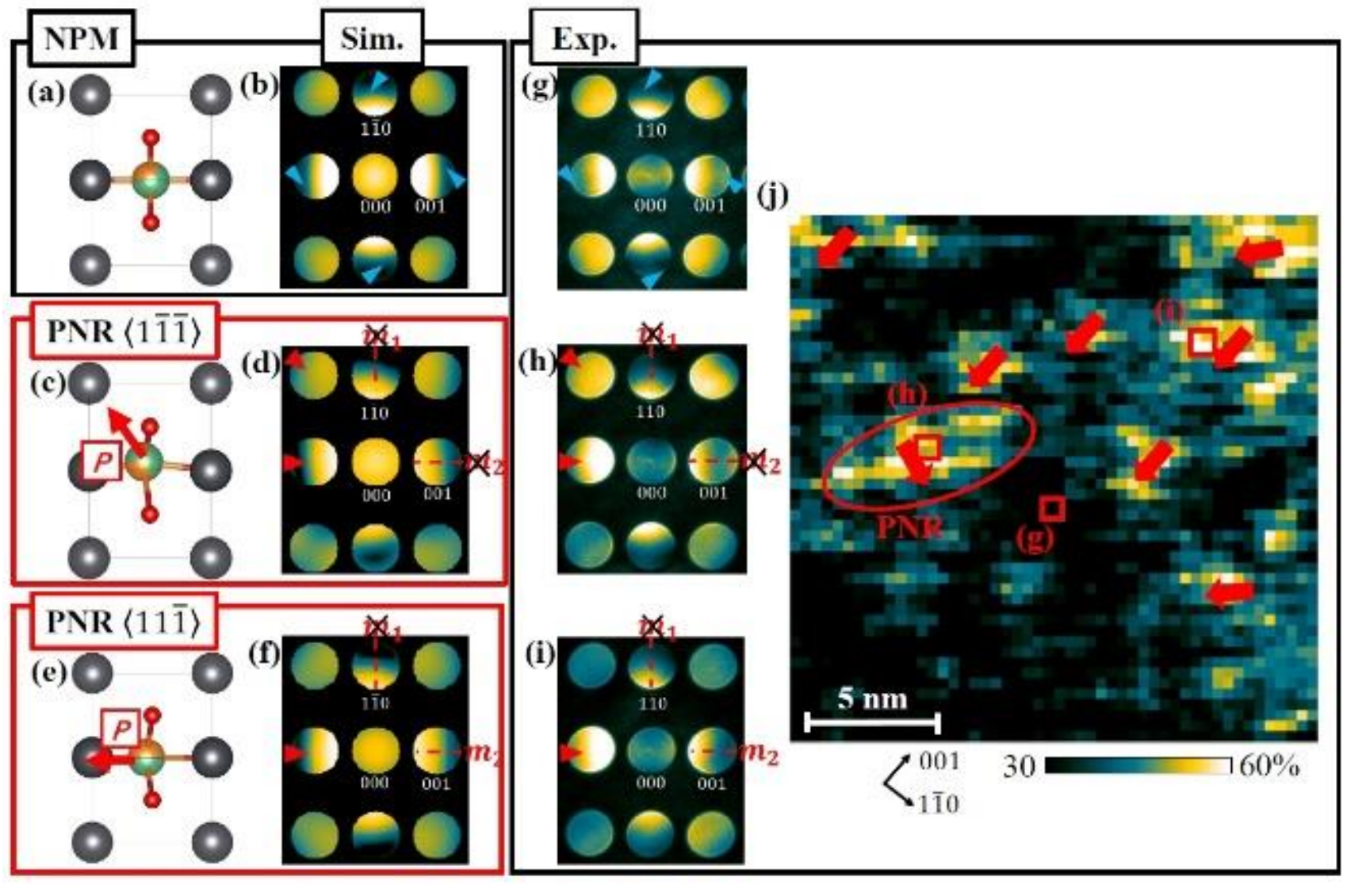

FIG 2. $[110]_c$ projected crystal structures of (a) the non-polar matrix (NPM), and polar nanoregions (PNRs) in (c) and (e). Corresponding simulated convergent-beam electron diffraction (CBED) patterns are shown in (b), (d), and (f), respectively. The NPM exhibits ***2mm*** symmetry, with both two-fold rotational and mirror symmetries clearly preserved. In contrast, the CBED patterns of PNRs show a breakdown of the two-fold rotational symmetry due to local polar distortions. Panels (g)–(i) display experimental CBED patterns acquired at the probe positions marked in the PNR map shown in (j). The PNR map in (j) was constructed by evaluating the symmetry-breaking index for two-fold rotational symmetry at each probe position, with a color scale from 30% to 60%. Bright regions correspond to PNRs. Red arrows indicate the inferred local polarization directions, determined by comparing experimental and simulated CBED patterns.

FIG 3. $[110]_c$ projected crystal structures of (a) the non-polar matrix (NPM) and (c) the chemically ordered region (COR). The corresponding simulated CBED patterns are shown in (b) and (d), respectively. In CORs, ordering of Mg and Nb at the *B*-sites forms a $2\times2\times2$ superlattice, giving rise to additional diffraction intensities—superlattice reflections—highlighted by a red circle in (d) as an example. Panels (e) and (f) display experimental CBED patterns taken at the probe positions marked in the COR map (g). The COR map in (g) was constructed by quantifying the superlattice reflection intensities across the 4D-STEM dataset. The color bar indicates intensity values ranging from 7.2 $e^6$ to 2.5 $e^7$, with brighter regions corresponding to stronger COR signatures.

FIG 4. (a) Real-space map of polar nanoregions (PNRs) at room temperature (295 K). (b) Corresponding map of chemically ordered regions (CORs) at 295 K. (c) Schematic illustration of the spatial distribution of PNRs (red) and CORs (blue) in the same region at room temperature. (d) PNR map at low temperature (100 K), showing growth and interconnection of PNRs. (e) COR map at 100 K, demonstrating the temperature-invariant nature of CORs. (f) Schematic representation of PNR and COR distributions at 100 K. The maps in (a), (b), (d), and (e) were obtained from the same sample area, enabling direct comparison across temperature. Minor vertical shifts in features are due to specimen drift and do not affect the interpretation.

References

[1] G. Burns and F. H. Dacol, *Solid State Commun.* **42**, 9 (1982).
[2] L. E. Cross, *Ferroelectrics* **151**, 305 (1994).
[3] S. E. Park and T. R. Shrout, *J. Appl. Phys.* **82**, 1804 (1997).
[4] S. J. Zhang and F. Li, *J. Appl. Phys.* **111**, 031301 (2012).
[5] A. D. Hilton, D. J. Barber, C. A. Randall, and T. R. Shrout, *J. Mater. Sci*. **25**, 3461 (1990).
[6] C. Boulesteix, F. Varnier, A. Llebaria, and E. Husson, *J. Solid State Chem.* **108**, 141 (1994).
[7] E. Husson, M. Chubb, and A. Morell, *Mater. Res. Bull.* **23**, 357 (1988).
[8] I. K. Jeong, T. W. Darling, J. K. Lee, T. Proffen, R. H. Heffner, J. S. Park, K. S. Hong, W. Dmowski, and T. Egami, *Phys. Rev. Lett.* **94**, 147602 (2005).
[9] D. Fu, H. Taniguchi, M. Itoh, and S. Mori, *Advances in Ferroelectrics.* InTech; 2012. Available from: http://dx.doi.org/10.5772/52139
[10] G. Xu, G. Shirane, J. R. D. Copley, and P. M. Gehring, *Phys. Rev.* B **69**, 064112 (2004).
[11] D. Fu, H. Taniguchi, M. Itoh, S. Y. Koshihara, N. Yamamoto, and S. Mori, *Phys. Rev. Lett*. **103**, 207601 (2009).
[12] E. V. Colla, E. Y. Koroleva, N. M. Okuneva, and S. B. Vakhrushev, *Phys. Rev. Lett.* **74**, 1681 (1995).
[13] A. Levstik, Z. Kutnjak, C. Filipic, and R. Pirc, *Phys. Rev*. B **57**, 11204 (1998).
[14] R. Pirc and R. Blinc, *Phys. Rev*. B **60**, 13470 (1999).
[15] V. Bobnar, Z. Kutnjak, R. Pirc, R. Blinc, and A. Levstik, *Phys. Rev. Lett.* **84**, 5892 (2000).
[16] V. V. Westphal, W. Kleemann, and M. D. Glinchuk, *Phys. Rev. Lett.* **68**, 847 (1992).
[17] R. Fisch, *Phys. Rev*. B **67**, 094110 (2003).
[18] Y. Moriya, H. Kawaji, T. Tojo, and T. Atake, *Phys. Rev. Lett.* **90**, 205901 (2003).
[19] B. P. Burton, E. Cockayne, S. Tinte, and U. V. Waghmare, *Phase Transitions* **79**, 91(2006).
[20] M. Pasciak, M. Wolcyrz, and A. Pietraszko, *Phys. Rev*. B **76**, 014117 (2007).
[21] A. Kumar, J. N. Baker, P. C. Bowes, M. J. Cabral, S. Zhang, E. C. Dickey, D. L. Irving, and J. M. LeBeau, *Nat. Mater.* **20**, 62 (2021).
[22] K. Tsuda, R. Sano, and M. Tanaka, *Phys. Rev.* B **86**, 214106 (2012).
[23] K. Tsuda, A. Yasuhara, and M. Tanaka, *Appl. Phys. Lett.* **103**, 082908 (2013).
[24] K. Tsuda and M. Tanaka, *Appl. Phys. Express* **9**, 071501 (2016).
[25] D. Morikawa, Y. Noguchi, and K. Tsuda, *Jpn. J. Appl. Phys.* **62**, SM1003 (2023).
[26] K. Tsuda and M. Tanaka, *Acta Crystallogr.* A **55**, 939 (1999).
[27] B. Dkhil, J. M. Kiat, G. Calvarin, G. Baldinozzi, S. B. Vakhrushev, and E. Suard, *Phys. Rev.* B **65**, 024104 (2001).
[28] D. Morikawa, M. Ageishi, K. Sato, K. Tsuda, and M. Terauchi, *Microscopy (Oxf.)* **70**, 394 (2021).
[29] T. Egami, W. Dmowski, S. Teslic, P. K. Davies, I. W. Chen & H. Chen, *Ferroelectrics* **206**, 231(1998).
[30] M. Akbas and P. Davies, *J. Am. Ceram. Soc.* **83**, 119 (2000).
[31] J. Chen, H. M. Chan, and M. Harmer,*J. Am. Ceram. Soc.* **72**, 593(1989).
[32] Y. Yan and S. J. Pennycook, Z. Xu and D. Viehland, *Appl. Phys. Lett.* **72**, 3145 (1998).
[33] M. J. Cabral, S. Zhang, E. C. Dickey, and J. M. LeBeau, *Appl. Phys. Lett.* 112, 082901 (2018).
[34] S. Wakimoto, C. Stock, R. J. Birgeneau, Z. G. Ye, W. Chen, W. J. L. Buyers, P. M. Gehring, and G. Shirane, *Phys. Rev*. B **65**, 172105 (2002).
[35] G. Xu, Z. Zhong, H. Hiraka, and G. Shirane, *Phys. Rev. B* **70**, 174109(2004).
[36] I.-W. Chen, P. Li, and Y. Wang, *J. Phys. Chem Solids* **57**, 1525 (1996).
[37] I.-W. Chen, *J. Phys. Chem Solids* **61**, 197-208 (2000).